# Holistic B2X Mobile Application Development – A Reference Model

**Oliver Werth** (corresponding author), Federal University of Applied Administrative Sciences, oliver.werth@bwz.bund.de, https://orcid.org/0000-0002-6767-5905

**Nadine Guhr**, OWL University of Applied Sciences and Arts, nadine.guhr@th-owl.de, https://orcid.org/0000-0001-8812-1488

**Michael H. Breitner**, Leibniz University Hannover, breitner@iwi.uni-hannover.de, https://orcid.org/0000-0001-7315-3022

Business-to-X (B2X, "X" for business, customer, etc.) mobile applications ("apps") show various mobile-specific chances and challenges that must be addressed in the whole development process. Knowledge about mobile app development models' usage remains restricted, although various process models for B2X app development have been published already. This article first reviewed available process models for mobile app development. In addition, 28 expert interviews with various stakeholders involved in typical B2X mobile app development processes were conducted to examine this research-practice knowledge divide. Since hybrid process models are often advantageous for B2X mobile app development, technical backgrounds or communication processes, are also crucial. Since no process model can be used unadopted, this study theorizes mobile-specific characteristics and challenges of app development process models to a reference model generally valuable for management decision support. The findings create avenues for better theorizing toward successful B2X mobile app development.

## Introduction and research needs

Mobile applications ("apps") and their development have received rapidly growing interest in recent years (e.g., 1–3). Today's apps are designed to be more user- and device-centric. Developers need to provide alternatives for different types of clients and users to customize their unique user experience (UX) [4, 5]. In particular, app development has several mobile device-specific characteristics and challenges, such as multi-operating system programming or sensor usage, and accessibility barriers which must be considered during the development process [6, 7]. Existing software engineering process models are often not directly applicable to app development [8]. In companies, process models must be implemented in a complex environment between the development team, plurality of project stakeholders, and information technology (IT) ecosystems (e.g., 9). The variety of process models, the ambiguity of correlations to concrete use cases, and the diverse domains of apps raise the question of which process models were implemented in practice and what influences the selection of a process model for their development [10].

With the awareness of these challenges, several process models for app development exist and have been discussed in the past (e.g., 3, 11–13), although the development process in practice is often quite ad hoc and unstructured. Jabangwe et al. [3] found that, in their review of 20 app development processes, the relevance and rigor of the examined process models are relatively low. Only two processes are useful in practice showing that process models in their status seem not helpful for practitioners. Currently, no structured approach is available in the literature that incorporates and systematically synthesizes mobile-specific app development knowledge to the process models. App development processes and their design methods are theoretically underdeveloped and lack in-depth research toward a reference model that can serve as a blueprint for developing this emergent class of socio-technical information systems (IS). Given the wide landscape of potential stakeholders involved in a development process, standardization practices toward reference models can help overcome the challenge, e.g., communications towards rapid and valid development processes [14, 15]. The existing literature, however, lacks such a reference model for mobile app development.

The contributions of this paper are twofold: We conducted and analyzed 28 semi-structured interviews with B2X app developers to examine the current state of the art in B2X mobile app development, reasons for (non-) usage of process models, and chances and challenges associated with their usage. Following Rowlands and Kautz [16], we define these developers as all individuals, e.g., project managers and creators who interact with the customer; in determining requirements, proposing solutions, and developing apps. We analyzed our interview transcripts using qualitative structured content-analytical methods according to Corbin and Strauss [17]. As we compared our qualitative research findings with the existing literature, we found potential pitfalls between research and practitioners. We third follow the comment of Rai [18, p. vi] that the "…interaction of observations with a variety of theories, models, and literature that provide alternative candidate abstractions can be used to spot breakdowns in viewpoints […] and can be a basis to generate a novel lens to the problem by bringing together complementary perspectives." Our proposed reference model serves as an accumulated knowledge foundation for more suitable and specific research, and it has high practical and theoretical relevance in B2X mobile app development. At the same time, it goes beyond questionable, already existing process models. The overall research objective is to accumulate our examinations and build a reference model to present a blueprint for B2X mobile app development projects for various interested stakeholders, e.g., project managers or project controllers. Consequently, this study aims to contribute to the state-of-the-art on substantial B2X mobile app development through current practice examinations, as proposed by, e.g., Jabangwe et al. [3] and Martinez et al. [19]. Furthermore, we assist practitioners to establish hardware and OS-independent guidelines and model understandability for B2X mobile app development as acknowledged, e.g., by El-Tom et al. [20].

In the following theoretical foundation, we provide the background of B2X mobile apps and process development models. Then, we explain and ground our research design and research methods. Subsequently, we discuss our results and deduce our findings, present our reference model and deduce implications and recommendations for theory and practice. Finally, we discuss limitations and outline further research directions.

## B2X mobile applications and development process models

We follow Hoehle and Venkatesh's [21, p. 437] definition of mobile apps as "an IT software artifact that is specifically developed for mobile operating systems (OS) installed on handheld devices, such as smartphones or tablet computers." Furthermore, we use the term B2X apps, implying mobile apps designed for business-to-business and business-to-customer contexts. Apps have numerous specific opportunities and challenges for users, app providers, and managers that must be considered in their development process. Numerous process models for B2X mobile app development exist. Taking a developer's perspective on the challenges, Joorabchi et al. [7] examined the existence of (1) general challenges, e.g., data intensity of the application, (2) challenges in multi-platform development (Android or iOS OS), and (3) testing. In addition, apps and their mobile values delivered are designed for specific purposes, and user groups must be considered in the entire development process [22]. In contrast, it is alarming that there is an increasing number of vulnerable, non-successful apps, even if the app market is facilitating the business [23].

Process models serve as structured guidelines that describe and order phases, tasks, and deliverables associated with app development. They describe events in a sequential arrangement that lead to an outcome [24]. Plus, they help involved people analyze and rearrange the activities that must be performed [25–27]. Given process models specifications for the software development domain, we follow the definition of Scacchi [28, p. 4] and define process models as a model to "represent a networked sequence of activities, objects, transformations, and events that embody strategies for accomplishing software," in this case, B2X mobile apps. Practitioners use and adapt (well-)known traditional software development process models to structure processes more efficiently. To the best of our knowledge, three research articles use a holistic view of this topic, focusing on process models for (mobile) app development [3, 11, 12].

Kemper and Wolff [12] are outdated in terms of such a vital topic of mobile apps, they focused on the appropriate selection of iterative process models for mobile app development. Jabangwe et al. [3] examined 20 different process models using a structured literature review. Their research on process models suggests that both rigor and relevance were low. Consequently, the usefulness of such process models is questionable. Werth et al. [11] also focused on process models, although not specifically designed for mobile app development, but with a holistic view of process models from software engineering suitable for app development. Using a taxonomic approach [29], they classified and structured 16 process models and methodologies for (mobile) software development. In the following, we extended this taxonomy and enrich it with a recent look in the literature (see Table 1).

Werth et al. [11] identified the waterfall model (WM) [30], spiral model (SM) [31], V-Model (VM) [32], incremental development model (IDM) [33], concurrent development model (CDM) [34], component-based software development model (CSDM) [34], extreme programming (XP) [35], SCRUM [36], dynamic systems development method (DSDM) [37], feature driven development (FDD) [38], the rational unified process (RUP) [39], Mobile-D [40], iterative process models for mobile applications (IPM) [8], lean software development (LSD) [41], MASAM methodology [42], and the perspective-dissolve approach for mobile application development (MobiPDA) [43].

We compared the process models of Jabangwe et al. [3] with the already investigated process models of Werth et al. [11]. Nine of these new process models did not provide a name, which we labeled PM1 [44], PM2 [45], PM3 [46], PM4 [47], PM5 [48], PM6 [49], PM7 [50], PM8 [51], and PM9 [52]. The other process models are the 3D-Approach, [53], AppSpec [54], InterMod methodology [55], Lean Agile Web Approach (LAWA) [56], Mobile Application Development Lifecycle (MADLC) [57], Magni [58], miSEL-sdp [59], Mobile Development Process Spiral [60], and Visual Environment for Designing Interactive Learning Scenarios (VEDILS) [61]. Furthermore, we found another process model ALP-mobile [62], through backward searches with Google Scholar. Of the 35 process models, 24 focused on mobile app development. We found that most (n = 27) of the analyzed process model and methodologies follow an iterative approach within the core dimension "phases." Only two follow a parallel approach (CDM and IDM); none of these is a mobile-specific process model. Most process models require at least a basic understanding of the clients/customers' requirements (n = 23), whereas twelve require a well-known understanding. A total of 31 process models used incremental prototyping in their process, whereas four did not provide any prototype to the stakeholders (PM3, PM8, VM, and WM). Two of them (PM3 and PM8) are mobile app development specific. Consequently, the inclusion of new objects in Table 1, leads to consistent findings with a clear tendency toward prototyping, compared to Werth et al. [11]. Taking a holistic view of all process models, 19 involve client/customers regularly in the development process (dimension "involvement of the client/customer"). The minority only includes them all the time in the software development process. Six process models (FDD, LAWA, LSD, MASAM, RUP, and XP) permanently involve the stakeholders in the development process. With a more detailed view of mobile app development models, we observe a clear tendency to involve the client/customer singular or regular (n = 21) but not on a permanent base (n = 3).

Table 1: Overview of existing process models and methodologies (roughly oriented on Werth et al. [11])

| Process models (alphabetical) | Core taxonomy dimensions | | | | | | | | | | | | | |
|---|---|---|---|---|---|---|---|---|---|---|---|---|---|---|
| | Mobile application specification | | Phases | | | Development efforts | | Requirements of clients/customers | | Prototyping | | Involvement of the client/customer | | |
| | Characteristics | | | | | | | | | | | | | |
| | Yes | No | Sequential | Iterative | Parallel | Low | High | Partially-known | Well-known | None | Incremental | Singular | Regular | Permanent |
| CDM | | x | | | x | x | | x | | | x | | x | |
| CSDM | | x | | x | | x | | | x | | x | x | | |
| DSDM | | x | | x | | x | | x | | | x | | x | |
| FDD | | x | x | | | | x | | x | | x | | | x |
| IDM | | x | | | x | | x | x | | | x | | x | |
| RUP | | x | | x | | | x | x | | | x | | | x |
| SCRUM | | x | | x | | | x | | x | | x | | x | |
| SM | | x | | x | | | x | x | | | x | | x | |
| VM | | x | | x | | | x | x | | x | | | x | |
| WM | | x | x | | | x | | | x | x | | x | | |
| XP | | x | | x | | x | | x | | | x | | | x |
| 3D | x | | x | | | x | | | x | | x | x | | |
| ALP-mobile | x | | | x | | x | | x | | | x | | x | |
| AppSpec | x | | | x | | | x | | x | | x | x | | |
| InterMod Methodology | x | | | x | | | x | x | | | x | | x | |
| IPM | x | | | x | | | x | x | | | x | | x | |
| LAWA | x | | | x | | | x | x | | | x | | | x |
| LSD | x | | | x | | | x | x | | | x | | | x |
| MADLC | x | | | x | | x | | x | | | x | | x | |
| Magni | x | | | x | | | x | x | | | x | | x | |
| MASAM | x | | | x | | | x | x | | | x | | | x |
| miSEL-sdp | x | | | x | | | x | x | | | x | | x | |
| Mobile Development Process Spiral | x | | | x | | x | | x | | | x | | x | |
| Mobile-D | x | | | x | | | x | x | | | x | | x | |
| MobiPDA | x | | | x | | x | | | x | | x | x | | |
| PM1 | x | | x | | | | x | | x | | x | | x | |
| PM2 | x | | | x | | | x | x | | | x | | x | |
| PM3 | x | | x | | | x | | | x | x | | x | | |
| PM4 | x | | | x | | | x | | x | | x | x | | |
| PM5 | x | | x | | | | x | x | | | x | x | | |
| PM6 | x | | | x | | | x | | x | | x | x | | |
| PM7 | x | | | x | | | x | x | | | x | | x | |
| PM8 | x | | | x | | x | | x | | x | | x | | |
| PM9 | x | | | x | | | x | x | | | x | | x | |
| VEDILS | x | | | x | | x | | | x | | x | | x | |

In result, we found numerous process models for (mobile) app development in the literature. Our taxonomy expansion from Werth et al. [11] leads to nine process models with the same specifics. InterMod methodology, IPM, Magni, miSEL-sdp, Mobile-D, PM2, PM7, PM9, and SM follow iterative phases, have a high development effort, require at least basic knowledge of the client or client/customer requirements, propose incremental prototyping, and involve stakeholders regularly in the software development process. Except for SM, all the models are specially designed for mobile app development. Against this overview from the literature, practitioners often do not efficiently use the process models presented in research [3]. Research on process models for app development is not widespread in practice. It is questionable and interesting why several process models for app development exist, whereas the relevance for practitioners is low. Against this backdrop, we want to push the discussion away from process models for mobile app development based on these arguments and theorize the knowledge about developing mobile apps to a more holistic level. We will propose a reference model for B2X mobile app development that goes beyond existing process models, lead to more in-depth theoretical explorations as well as knowledge foundations, and serve as a blueprint for further examinations and development projects [14]. Also, our reference model provides a standardized framework that can be used across different organizations and mobile app-specific industries. It can establish common terminologies and structures, promoting consistency and serve as a general overview for organizations, reducing time and effort required to design and implement processes for their tailored purposes [63].

## Research design and methods

A two-step approach drives the research design—we first review process models for (mobile) app development with a taxonomic approach. In line with Dey [64], our initial examination of the relevant literature on process models for app development (see Table 1) presented within a taxonomic structure serves as the knowledge foundation for interview situations [65].

Second, we performed qualitative expert interviews and accumulated all information into a holistic reference model. Given the dynamic nature of B2X apps and development, we performed a qualitative approach with expert interviews. We follow the basic assumption of Gioia that "the organizational world is essentially socially constructed, […]" and that the "[…] informants are "knowledgeable agents" [66, p. 291]. This implies that B2X app developers in organizations "[…] know what they are trying to do and can explain to us quite knowledgeably what their thoughts, emotions, intentions, and actions are." [66, p. 291]. This allows us to generate new insights through direct interaction with interviewees and to obtain information on whether and to what extent the available process models from theory are also used in practice and whether "shadow models" dominate reality. To understand identity needs, the research approach we followed is an interpretative paradigm that regards participants as knowledgeable stakeholders and seeks to capture and model participants' experiences and meanings [67, 68]. To grasp the complex nature of B2X app development, we rely on a qualitative research design, often referred to as the "Gioia Methodology," following a qualitative-inductive approach [68]. This approach, according to Sarker et al. [65], requires principles of theoretical openness and methodological flexibility. It allows an inductive examination of insufficiently investigated phenomena [66, 69]. We aim for reference model elaboration, drawing on and extending important ideas from research and practitioners on app development. When preexisting conceptual ideas drive a study's design, theory elaboration eliminates the need for theory generation through purely inductive, grounded analysis [70, 71].

To gain valuable insights from practice and critically reflect on the process models available in theory, we followed a theoretical sampling strategy based on relevance, expertise, and appropriateness rather than representativeness [72]. Qualitative methods are well suited and valuable for investigating dynamic organizational processes [71], as they are sensitive to the organizational context and their potential for focusing upon activity sequences as they unfold. From our research objective, we developed a semi-structured interview guideline to conduct the interviews with an appropriate level of structure and sufficiently reliable results [73]. Hence, the guideline only acted as a rough outline, affording the interviewees and researchers considerable liberty. We performed a pretest to validate the guideline with students and professors, postdocs, and graduate students involved in the project. Subsequently, the interview guideline was adapted based on these comments. The final interview guideline consisted of five interview blocks. First (Block 1), a quick introduction and relevant information for the experts were provided, including background information on the research project, the domain, e.g., personal B2X portal applications, information for the upcoming process, and privacy information. This information enabled interviewees to evaluate whether they could provide expert knowledge of app development processes and their underlying models. The interviewees were also encouraged to speak openly about their biographical and software development experiences. In Block 2, questions were asked regarding process models in general, followed by questions about the usage of process models in B2X mobile app development (Block 3). In Block 4, questions were asked about using process models for B2X mobile app development in the domain the interviewee was connected with, e.g., personal finance. Two concluding questions were asked to provide feedback on the interview (Block 5). Hence, the interview guide was used to highlight potentially interesting topics or aspects in the conversation that were briefly touched upon by the interviewees and deemed relevant to the research question. The interviews provided in-depth accounts of B2X app development in practice app developers' experiences in a narrative form. The interview guideline can be found in the appendix of this article (see Appendix A.1).

Interview participants (Table 2) were invited via e-mail, personal contact, or telephone by the academic and practice network of the researcher's institutions. The sampling for interview partners intended to receive an overall opinion and impression about the process and the underlying process models for app development. We looked for interview partners with professional experience in B2X app development (projects) regardless of their development experience within a specific app category to receive a general view and expressions about our research topic. Consequentially, we assume that those interview partners are suitable to provide us information to fulfil our research aim. During the interviews, we adopted the role of neutral observers, although we were aware that this did not imply that we were unbiased. We obtained answers from different perspectives that were as straightforward as possible [74]. We gave the participants much room in the interviews to express their views on app development. The interviews were conducted between May 2019 and July 2020. Depending on the interview partner, they were held in German or English. The interviews lasted between 30 and 120 minutes. We used the criterion of theoretical saturation as the closing point for the collection of new data. Theoretical saturation is the point at which no new topics or insights are generated [75, 76]. After 28 interview situations with 29 experts, we found no new concepts, ideas, or topics, and thus we stopped additional interviews.

Table 2: Interview overview (sorted after interviewee profile)

| **Expert(s)** | **Profile of Interviewee – Role in Development Process** | **Experience in the Mobile App Category (oriented on Han et al. 2015)** |
|---|---|---|
| E1 | Agile Coach / Team Member | Personal Finance |
| E2 | Agile Coach / Team Member | Personal Finance |
| E3a / E3b | App Developer (a) / Team Member (b) | Personal Finance |
| E4 | App Developer | Personal Finance |
| E5 | App Developer | Personal Finance |
| E6 | App Developer | Personal Finance |
| E7 | App Developer | Lifestyle / E-Learning |
| E8 | App Developer | Lifestyle / E-Learning |
| E9 | App Developer | Lifestyle / E-Learning |
| E10 | App Developer | Personal Finance |
| E11 | App Developer | Personal Finance |
| E12 | App Developer | Lifestyle / Mobile Ticketing |
| E13 | Implementation Lead | Portal Applications |
| E14 | Lead Development Team | Personal Finance |
| E15 | Lead Development Team | Personal Finance |
| E16 | Lead Development Team | Personal Finance |
| E17 | Lead Development Team | Lifestyle / E-Learning |
| E18 | Lead Development Team | Portal Applications |
| E19 | Lead Development Team | Portal Applications |
| E20 | Lead Development Team | Utility / Mobility |
| E21 | Lead Development Team | Portal Search / Portal Applications |
| E22 | Lead Development Team | Lifestyle / E-Commerce |
| E23 | Product Owner | Personal Finance |
| E24 | Product Owner | Utility / Mobility |
| E25 | Project Manager / Team Member | Lifestyle / E-Learning |
| E26 | Scrum Master | Lifestyle / E-Commerce |
| E27 | Scrum Master | Personal Finance |
| E28 | Scrum Master | Utility / Mobility |

The data collection and analysis were tightly interwoven. Subsequently, we transcribed all interviews and analyzed the transcriptions using a qualitative structured content-analytical method, according to Corbin and Strauss [17]. The interview data set was organized, sorted, and centrally managed using MAXQDA Analytics Pro 2022. The coding procedure is a process that "gets the analyst off the empirical level by fracturing the data, then conceptually grouping it into codes that then become the theory which explains what is happening with the data" [77, p. 55] occurring in iterative steps. We selected this approach because it has the advantage of data reduction. It is flexible and systematic [17].

Furthermore, these techniques are advantageous for developing prescriptive knowledge about a topic to transform it into a reference model with generic topics in the second stage of our research design (e.g., 78). In addition to the advantages of the method, the following argument supports our decision: the research field considered here, with its complexity and contextuality and the associated diverse domains and phenomena that arise from related disciplines and concepts, needs an open, qualitative approach. The findings of these studies allow triangulation of the results [79] indicating that the data are based on different sources, e.g., interviews, information from literature, and white paper reports or guidelines about B2X app development.

We initiated data analysis after conducting our first expert interviews by carefully reading the interview material to understand the participants' activities and experiences. This provided a basis for inductively forming codes based on how the interview partners expressed themselves. During the ad hoc coding process of the data material, we modified the categories (both emergent and predefined) wherever necessary. After the first rounds of coding, the 13 first-order codes, e.g., statements about the usage of Scrum for B2X app development and corresponding anchor examples from our interview transcripts, were analyzed for their similarities and differences. Subsequently, the first-order codes were theorized. We compared, contrasted, and discussed the codes concerning concepts and ideas from the literature, leading to the identification of second-order themes, such as Scrum or adaptions of Scrum. Comparing and clustering these 13 second-order themes, we identified six aggregated dimensions, i.e., "selection of a specific process model," "hybridization and adaption of process models," "requirements engineering for mobile apps," "team characteristics and roles in development processes," "communication and information flow in development processes," and "governance issues and management of development processes." We sorted them into four main layers within the proposed reference model. An overview of first-order codes, anchor examples, second-order themes, and aggregated dimensions can be found in the appendix of this article; see Table A.2.

## A holistic view on B2X mobile application development – Insights from practitioners and towards a reference model

### Governance and IT project controlling – The management layer

During the interviews, it became apparent that the management must ensure several tasks and responsibilities have been subsumed under a management layer. A general statement was made regarding understanding process models: Some interviewees (E7 and E28) opined that the app development teams did not really understand the process models. In most cases, a company's management requests the development teams to use agile methods because several other companies also use them. As B2X app developer with more than six years of experience in the field (E16) from the financial services sector reported, transitioning from an agile process world to a waterfall model is difficult because of the inertia of internal employees:

*"So, if you come from the dynamic, agile Scrum world and then you start with this black and white waterfall model, so without building a bridge [for employees]. I think that doesn't work."*

Several employees do not understand the need to rethink or change established methods because of their perceived low usefulness. A changeover of all company stakeholders to agile process models is often desirable but not practicable. Several experts experienced that requirements were changed at short notice and had to be implemented during an active Scrum sprint because external stakeholders did not know about the sprints and their characteristics. Based on these statements, all development team members must understand the roles, processes, activities, and deliverables of a selected process model and be committed to the development project goals. The management should ideally ensure this. Estimates of time and budget by management are not always performed when using the agile model. This is in line with interviewee E5, where the development team remains committed to four releases per year, leading to unnecessary time pressure and planning difficulties. Scrum has disadvantages that occur in practice, for example, sprint duration. In some cases, company requirements regarding the B2X mobile app can change at short notice, e.g., quicker than two weeks, and Scrum does not provide for the objective of the sprint to be changed. An app developer (E8) acknowledges that:

*"Yes, we also use Scrum. The problem with agile methods for the development of apps is that it is very short-lived. There are many updates coming, system-side updates from Google or from Samsung [...], then the apps simply do not work anymore. It is difficult to make a risk and performance specification because when an update comes, for example, a new Android version, you need three or four months to make an update for the app, then the entire storage system no longer works."*

According to other interview partners (E2, E5, and E27), the management and clients of B2X apps are often still accustomed to classic process models and require an estimation of time and money expenditure at the beginning of the project. However, these experts mentioned that this is more difficult to perform with agile methods than with classical methods because neither the requirements are fixed at the beginning nor the number of required scrum sprints is known. These experts suggested an inverse estimation method. The client must not determine the budget alone—the developers attempt to create the best possible results from this budget. By frequently presenting the results after each Scrum sprint, the client can monitor the application's progress at any time and adjust the cash flow accordingly. Development team member E2 highlighted agile process models' general risk management problem. Because of the iterative character of these models, developers only often plan from sprint to sprint and do not have a long-term perspective.

Consequently, acute risks can be easily overlooked. In addition, some developers see risk management solely within the role of the Scrum master. However, according to expert E2, risk management belongs to the entire team and not exclusively to the management on the C-level.

Another team leader (E19) stated that the scope of a development project is essential for the degree of structure in the process provided by the central instance of the company. Small projects can be executed by an IT expert, more or less like its suites' needs, leading to mixed methods. Scrum is deemed a framework not implemented or tracked by official compliance. Expert E21 stated that integrating a scrum process into the IT department is not a problem. However, the remainder of the company, acting as stakeholders in the development process, are pushed to do so owing to the abundant culture and mentality of the company. Several people in positions of responsibility misjudge that implementing agile approaches does not only consist of learning and applying tools and techniques. Agility requires serious adaptations in a company's culture to work appropriately. A company's management culture must adapt to agility requirements. To achieve consistent understanding, e.g., regarding cost, development time, and B2X app requirements, the management must discuss and align a company's management culture and principles. For example, people's ideas and methods are more important than the subsequent processes. The associated release of time and creative energy will not succeed; however, if actual decisions are made according to the strict guidelines of the project management manual, creative lateral thinkers have little chance of asserting themselves but endanger their careers. Being creative in app development is important. Several experts state that sometimes the developers are not allowed to have this creativity, thereby limiting their agility.

Furthermore, agility indicates openness toward learning processes and taking the insights of a project seriously, according to Scrum master E28. E18 (lead development team) provided an example in which developers were taken out of their local teams and given directly to the client because the company could not deliver the fast iteration steps in the app development that the Scrum model promised. It was shown that agile methods work quite well in the isolated area of the development of one app. However, it becomes much more difficult when the parallelism of up to ten development projects is implemented by one development team. This leads to much more variability in the backlog than expected, and the backlog is always exhausted with future tasks. The company's management of a development team member (E25) acted excessively conservatively with developers' newly gained flexibility and freedom in agile environments. This implies that much of the agility and trust of the developers were lost in the process. As companies often have strict deadlines, they put pressure on the developer team in stressful situations. Developers drift away from these deadlines and quickly deviate from agile principles. Company management often limits the agility of the development team (E7). A scrum master (E28) stated that:

*"If stress or chaos comes, [...] still fall back into old thinking patterns and think again, one can only do this with classical methods, thus with many status reports and the management controls. More deadlines, more controlling that the developer has to write down every hour what they have done now."*

An agile B2X app development process must be accepted by the project management, development team, and clients/customers.

**Requirements engineering in B2X application development – The requirements layer**

As described in the theoretical backgrounds, depending on the nature of mobile apps and the corresponding technical backbone, i.e., the smartphone, several considerations about requirements arise that can be accumulated within a requirements layer. A development team lead (E17) claimed for a completely free app development, which is not bounded to a development process—owing to most of the requirements in the mobile app market, acting much faster than even an agile development process is necessary. If an idea is approved by most of the development team, it is implemented. This only works if the company can act fully autonomously in the definition of app functionalities. The context of the company matters, according to app developer E21. The B2B mobile app mentioned in the interview was the first of its kind for the company's employees.

Further, the expert (E21) stated that most of the users, i.e., the employees, had never had a smartphone in hand before. In such cases, the development process, based on agile methods, needs to be adapted to the context in which it is implemented. In practice, urgent requirements change or interrupt any active sprint during development. To handle rapidly changing requests, as many requirements as possible must be identified at the beginning of the development project. In contrast, the end-user of the B2X app is consulted less frequently. Here, experts E24, E25, and E26 still identify a need for improvement as end-users' feedback in the entire development process is generally considered valuable. Several interviewees said that the agile approach allows feedback to be provided promptly, both from the user and development team. This allows mistakes to be identified and corrected more quickly because agile methods only work on small tasks consecutively [52, 80]. Certain quality requirements are particularly important in B2X mobile app development, e.g., graphical user interfaces (GUI) and UX. The development of apps is strongly surface-driven. The conception and design of an app always focus on usability. The concept of UX emphasizes the orientation toward the user and calls for a purely functional approach to software applications to be supplemented and expanded. Emotional factors regarding designs and esthetics are increasingly considered, which can influence and ease the pleasure while using the app, according to app developers E7 and E8. The app must be designed so that a broad circle of users, e.g., young or elderly users, can use and understand it. Several experts mentioned that the testing effort for B2X mobile apps increased compared to desktop apps. One reason is the complexity of the GUI and its usability. According to app developer E5, it is advisable to start testing on the original hardware at an early stage:

*"When developing mobile applications, I would say that the most important thing is to start testing even faster on the device itself, not on any simulators or emulators, and really on the mobile device because usability is everything. The display is manageable, and the fingers are so and so thick. And it doesn't help me to turn up the resolution, as I might do on the computer."*

Design errors that cannot always be reproduced on software emulators, such as readability or performance, were avoided. In addition, an app's special characteristics for each mobile OS and the mechanism through which it securely handles app data must be carefully considered. Overall, we identified that domain-specific and mobile user requirements for B2X mobile app development should ideally be identified early in the development process and constantly evaluated with prototypes, including addressed hardware. As a result, we argue that the requirements layer is strongly connected to the communication layer presented later.

**Selection of process models and adaptations – The process layer**

Based on our motivations, we present the observations from the interviews regarding the (non-)usage of specific process models for BX2 mobile app development. We subsumed the results within the process layer of our reference model. Nearly all process models from the literature except established models, such as Scrum, XP, SM, and WM, were unknown to most of the interview partners. The interviews with the experts revealed that no process model was completely suitable for all requirements and app developments. However, with Scrum, one process model was mentioned most frequently by the interviewees and is usually preferable and most common for a B2X mobile app development process if all roles can be appointed properly. During the examination of previous research and the interview situations, the Scrum model is preferred because of more communication and structure-enhancing methods, but this only applies if a product owner (PO) is appointed. According to several interviewees, the PO must be included in the agile approach if the team is large enough and all positions are filled. This is often not possible in smaller companies because they do not have several employees who can fill all positions. In practice, the experts reveal that the complexity of a situation is mainly influenced by requirements, guidelines, and external stakeholders in app development. One main criticism of the interviewees is role allocation in the Scrum model. Some interviewees omitted the PO. Because no person was left to cater to a client's interests, the requirements cannot be sorted according to urgency or importance. Without a PO, the development team cannot decide whether the app's function is correctly and completely implemented. In practice, a PO with enough authorization and decision-making powers has enabled minimal waiting times for permissions and more easily enforcing agreements. In addition, because of their role as representatives of clients' opinions, they often do not belong to the development team but to the client, as expert E23 mentioned:

*"So, in the end, there is also a PO on the client side, but in principle, he does more stakeholder management for the client and is the interface to me. In other words, he collects requirements in-house, passes them on to me, and I then turn them into stories and make sure that they end up in the app, e.g., functions regarding UX."*

Hence, prioritizing requirements that are not the client's focus, e.g., the stability or UX of the app, is possible. In Scrum, the omission of the PO must be avoided. The same applies to the Scrum master. Their tasks include compliance with company rules and process model actions. Without such a role, it will inevitably lead to the breaking of rules, such as the cancelation of team meetings, according to expert E2. An implementation lead (E13) described Scrum's independent and self-responsible processes as challenging for the development team. The construct of stakeholders involved in the development process leads to ad hoc new prioritization in the product backlog at sprint changes or inside a running sprint. A skilled (more than 20 years of work experience) development team leader (E20) stated that one risk of a parallelized Scrum process is losing track of long-term goals because the focus can shift to small tasks queued in the backlog every day. It can be concluded that PO in B2X mobile app development teams are mandatory because they serve as client representatives and represent clients' requirements.

Classic sequential models, such as WM, are suitable for B2X mobile app projects with predefined requirements. This is typically the case if a company's management wants to calculate the budget, time, and work required to complete the project as accurately as possible. For example, financial institutions are affected by several regulatory requirements. Because apps must also fulfill several regulatory requirements in the personal finance sector, classic models, e.g., WM, can be advantageous. Owing to the special characteristics of personal finance apps, interviewee E28 suggests that the WM is an appropriate model for app development. In the case of finance apps, security requirements or the usage of highly personal information arise [81]. Repetitive tasks (i.e., already known requirements fundamentally from the app environment to developers) can be performed more easily. Scrum Master E28 states that:

*"[...] The agile process is simply an empirical process. It assumes that it is important to create good software and that a learning process is also crucial in the development of an app. And that's why agile methods place so much emphasis on regular feedback, on making the app testable at an early stage, and on making it possible for customers to really experience it so that they can get feedback at an early stage as to whether they are moving in the right direction or not. [However] if it is in a complicated area or you know exactly what to do, like in a simple software migration, or you need a product after, the client wants the same product, [i.e., mobile app] as before, but then you can also go super on classical methods [or models]."*

Other processes, such as VM, XP, and SM have low relevance. The reasons for this are lack of communication with the client, inflexibility regarding changing and unknown requirements, and slow detection of problems or errors (testing), which are necessary for the interactive character of B2X mobile app development. A development team member (E2) acknowledges that agile models are generally more effective in implementing short-notice guidelines and working environments with small teams than plan-driven models like the WM. This is also underlined by Mousaei & Gandomani [80], who stated that agile development is more appropriate due to shorter responding time to changes and significant ways of speeding up the development of mobile apps. Classical models often fall short. A strictly linear division into clearly distinguishable project phases, in which a certain goal is always achieved before the beginning of the next phase, is often impossible in app development. However, our findings show that classic models are usually preferable for B2X mobile app development with fully pre-known general, mobile-and domain-specific requirements. Based on our dataset we can claim that this is the case for the high-regulated financial services sector without being exhaustive to other business sectors.

Most interviewees agreed that simply adapting well-known models from the literature is impossible. Adaptation and hybridization are combinations of models that are often necessary. An app developer (E10) states that sprint length, team size, and composition must be adapted. In Scrum, sprint lengths of more than two weeks cannot be maintained efficiently. Several companies need a longer sprint length to release the B2X app. Scrum master (E28) argues that inexperienced teams must adhere to strict guidelines because numerous mistakes can be made, particularly at the beginning of the development process.

Conversely, experienced teams can make deviations from the models that create added value. A Kanban board can add structure to the sprint phases according to the statements of a development team member (E2) and a Scrum master (E26). The combination of the two models allows processes from external teams and policies from supervisory authorities to be integrated more easily. Two other experts can confirm this, i.e., E10 and E11, for mobile app development in the financial services sector: These experts proposed Scrum as a methodology combined with WM for their B2X mobile app domain. A sequential process is advantageous if several requirements are clear before the development process, e.g., security aspects of the app.

Furthermore, other stakeholders like data centers want constantly changing requirements. Short-term feedback that can be obtained through the Scrum process is essential, and it can identify possible undesirable developments at an early stage. Generally, as can also be observed from expert E27, the process models can only be understood as a basic framework. Guidelines must be individually adapted to the company for the largest possible benefit. However, the practice has shown that several changes are not advantageous. An app developer (E10) marked the following:

*"I think you always have to think about how much I take a standard process and how much I have to adapt this standard process to reality if in doubt. I don't think you can always adopt a template from the textbook and say that it works for everyone else; let's take that as well."*

The interviews show that hybridizations and adaption of process models for B2X app development can be advantageous if recurrent requirements are well known before and during a project.

**Information flows in B2X application development– The communication layer**

Internal and external communication are essential components of B2X mobile app development process models that are included in the communication layer provided. However, it can be observed in practice that the developers in the development team often do not exhibit the necessary motivation to communicate. According to the interviewees, this is mainly because developers consider programming and implementing as their main tasks in the team and perceive all obstacles deviating from this task as unnecessary. Similarly, developers' perspectives and programming experience must be considered. According to PO (E24), developers' feedback is extremely positive for the entire development process. They must be in close contact with the PO to prioritize the requirements in the Scrum backlog more effectively. Several experts consider the daily Scrum important, exceeding the proposed length of 15 min. Here, the daily Scrum was excessively long, did not provide enough new information, and often addressed requirements that were not always relevant to the involved. According to an app developer from the personal finance domain (E6), the main topics of the discussion are obstacles that affect other team members and provide an overview of the status. Any information irrelevant to all people present must be discussed separately with the respective employee. These proposals enable a more efficient daily Scrum and sprint planning.

Several process models in past improve communication with clients or stakeholders. Experts mentioned that requested communication with clients regarding future planning, often several Scrum sprints in advance, is a general disadvantage of the Scrum model. It is often necessary to disclose further information to the client whose features will be available, even if this is not completely in line with the agile model. An important point is transparency in the development process. An app developer (E10) reveals that it is always important to be transparent with the client and to provide interim results about the development. Through small, short iterations, the developers can always provide the client with intermediate results and a better product overview. Another challenge developers face is implementing the design laid out by GUI designers. According to some interviewees (E7, E8, and E9), developers sometimes do not understand the requested GUI design and have problems implementing it. Most of the time, B2X mobile app developers are not involved from the beginning and cannot implement anything from the designers. An interviewee (E8) says that developers must state their opinions about the design in the beginning:

*"That the programmers are there right from the start when new features or functions are virtually invented. [...] But at least one with technical understanding is there and can say what works and what doesn't work. [...] And that is why we have now also found a product discovery process where the developers themselves are involved in the discovery of new features. That means we are there right from the start. [For example], the company wants to have a new map feature within the app [...] we are there from the beginning and can say what works for the users and what is technically possible."*

We conclude that a high user-centricity in a B2X mobile app development process makes permanent and efficient internal and external project communication between team members and all relevant stakeholders mandatory and causes a challenging learning process for all.

**A reference model for holistic B2X mobile application development (REMOB)**

While the results from our literature review provided significant academic findings, the reference model for holistic B2X mobile application development (REMOB) displayed guidelines in more suitable format so that they could be used by mobile app developers. The interviews reveal that a holistic view of B2X mobile app development can be subsumed into four main layers associated with tasks, issues, and special importance on B2X mobile app development (indicated through a smartphone pictogram). Dashed borders indicate tasks and issues that arise if agility is used for B2X mobile app development. We integrate the findings into a reference model for mobile B2X app development (REMOB, see Figure 1).

While the management layer deals with more general issues on the whole development project, we conclude that the understanding of process models used by the team members and risk management are generally true for software development. Both are not mobile-specific but are certainly important for B2X mobile app development projects. The agility used frequently for B2X mobile app development is constructed upon the lean idea, affecting a separate department and representing a new way of understanding work processes and possibilities to improve daily work [36, 82]. Management must implement these principles in all intersecting departments and for all stakeholders in a B2X app development process. App developer E7 explained that several companies use given process models because the management forces them. App developer E10 switched from WM to Scrum because Scrum is used in practice by other firms, and they wanted to try it as well, even if the development could still be implemented with WM. The duality of management mentality and culture leads to extra work and extra effort for the development team and must be constantly discussed. Because of the strong dominance of agile work for B2X mobile app development, we conclude that these statements are of special importance for the mobile domain.

The requirements layer is concerned with more technical aspects of B2X mobile apps and the appropriate tests of these aspects. Intuitively they plan important, mobile-specific importance in the development process. Here, specific requirements such as GUI design arise because apps are developed for handheld devices. In development processes, identification of these requirements and continuous testing in the final environment, with the OS and hardware, is strongly recommended, as OS is one of the main critical success factors for value creation in a mobile context [83].

The process layer demonstrates the diversity of used process models and must be carefully considered by interested stakeholders of REMOB. Previous research suggests that classical process models lack flexibility in a company environment [13, 27]. An originally planned development process must often be changed because of unintended scenarios, e.g., changes in time or budget restrictions [84]. This is particularly true for app development processes [11, 12]. The high volatile requirements of mobile B2X apps require a more plan-driven adaptive development approach. At the same time, agility is ideal for app development projects with high levels of variability or uncertainty [80].

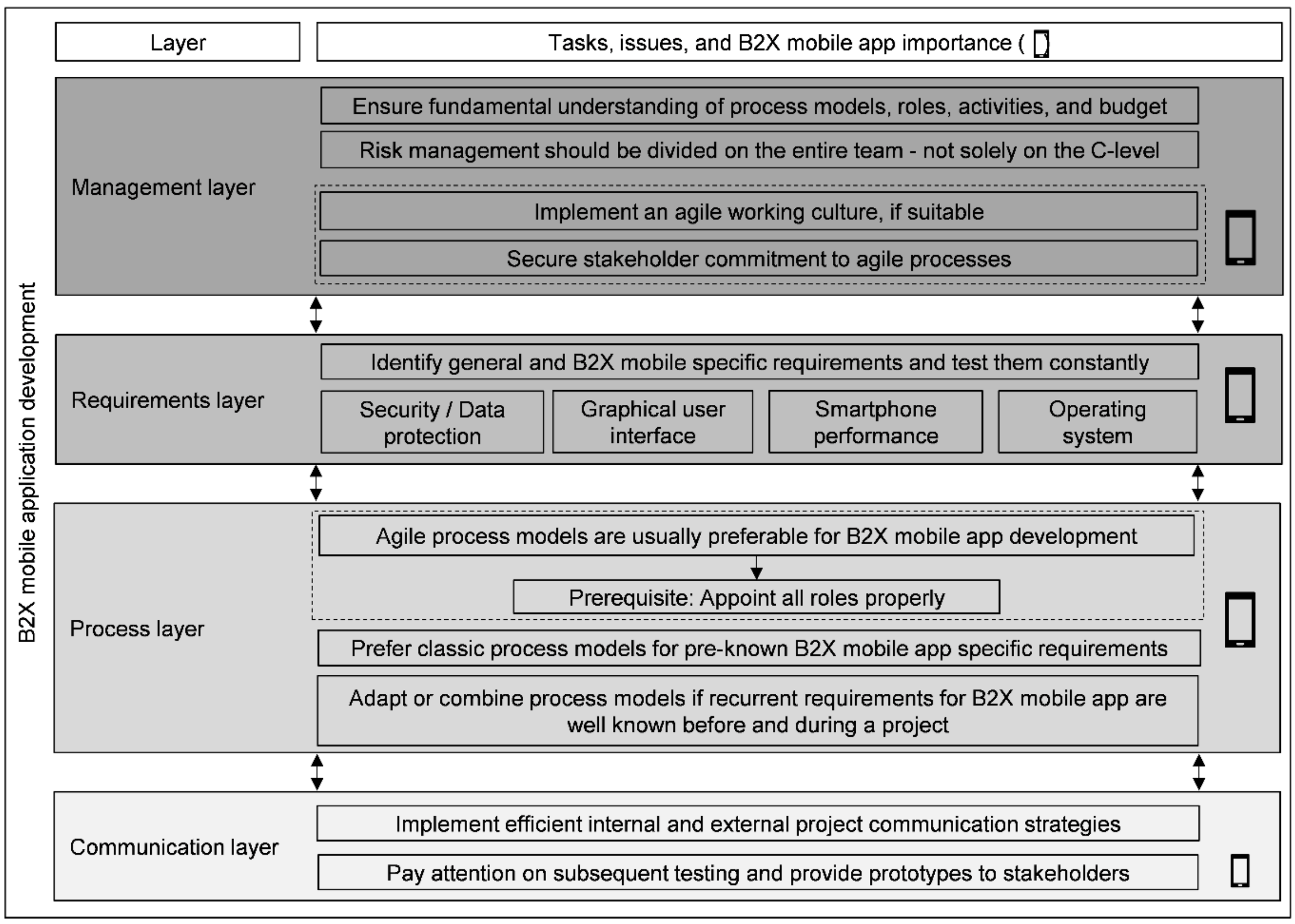

Figure 1: The reference model for holistic B2X mobile application development (REMOB)

To some extent, agile software development methods are appropriate for B2X mobile app development because they guide information flows more easily. It can be better tailored to unintended scenarios or company environments. Based on our expert interviews, companies often attempt to implement agile approaches, such as Scrum. Our interviews showed the important role of PO in the development process. Because of the high user-centricity and usability of apps (e.g., 21), a client representative (i.e., the PO) is mandatory. According to previous research, classical methods are often not relevant to the development of apps. Changes during the project are complicated and time-consuming and can lead to considerable delays, as all plans must be carefully updated.

Nonetheless, our findings have demonstrated that classical methods can be appropriate, particularly for pre-known B2X mobile app-specific requirements, such as security requirements in personal mobile finance apps. However, conventional approaches still have the authority to exist. In this case, it is also recommended to continue using conventional methods and combine them with agile methods when such requirements are recurrent.

Lastly, we identified a communication layer that deals with an appropriate communication strategy by all stakeholders. However, this statement can be true for all software development projects. Viewing the communication layer, user-alignment is reflected in the continuous and efficient communication between internal and external stakeholders. An efficient testing strategy must be employed and conducted within the B2X mobile app development process.

## Discussion, implications, and recommendations

In this research, we first reviewed the literature on process models for general software engineering and app development. We conclude that the table, i.e., Table 1, in a taxonomic structure, is useful for structuring the research field in a transparent, non-ad-hoc manner. This served as an appropriate knowledge foundation for the interview situations. Evaluating such artifacts, i.e., taxonomies, is an important step toward their usefulness [85, 86]. In addition, the core taxonomy dimensions were useful as an ignition to identify first-order codes and second-order themes, e.g., testing in the interview materials. Furthermore, the taxonomy plus the aggregated dimensions derived from the qualitative interview analysis served as a starting point toward a holistic reference model that serves as a blueprint for management and responsible persons towards efficient B2X mobile development projects. To the best on our knowledge, we did not find this connection and implication in the literature before, leaving this as an interesting observation for researchers. We implicate that taxonomies serve as an eloquent starting point for model-building intentions, i.e., reference models. In the case of B2X mobile app development, however, additional qualitative data from the field was needed to grasp the associated domain- and mobile-specific issues. A comparison of the core dimensions in Table 1 with the aggregated dimensions derived from the interviews reveals that new dimensions were identified that were not included in the taxonomy based on literature, e.g., "team characteristics and roles in development processes." Further data collection from the field was necessary.

In result, we propose REMOB, a reference model for B2X mobile app development that incorporates and describes the specifics of B2X mobile app development in a holistic manner. By implication, our collection of interviews proved that agile methods, such as Scrum, are widely used for app development. This is consistent with previous literature (e.g., 19). Several other process models from the literature are widely unknown and have also been proven by some of our interviewees. Management and interested stakeholders that plan a B2X mobile app development project must be aware of tasks and issues within four main layers. We reflect and expand communities' knowledge of mobile-specific process models in an exploratory manner and provide REMOB. We investigated the research-rigor divide for mobile application process models, as examined by Jabangwe et al. [3]. Our research closes this divide for further research as a theoretical implication, as proposed by Jabangwe et al. [3] and Martínez et al. [19]. It examines how and why existing process models for app development are used or not. In general, more process models for B2X mobile app development are not useful for practitioners because the existing models are already widely unknown. We proposed REMOB, taking a holistic perspective on the development process and generating a novel lens to the problem by bringing together complementary perspectives [18] to close better the gap between research outcomes and practitioners' interests and needs. The finding supports that developers and project leaders or project controllers adapt, change, or roughly orient themselves on (well-known) process models. Using REMOB organizations can enhance development efficiency by orienting themselves on a general set of layers, tasks and issues that are (or not directly) connected to mobile

B2X app development.. Developers can leverage these resources to expedite the development process. Also, novice developers can benefit from REMOB as the first orientation for their projects if they are not forced to use given process models by their company. App developers can also use REMOB for their projects to efficiently develop successful apps for different domains. Failure causes and risks of software development projects have been extensively discussed in the literature, and several critical success factors have been examined (e.g., 13, 87), but they are not mobile-specific. We support a generalized discussion and knowledge platform with REMOB and the provided app development process models in the second section. This helps to reduce failures caused by a lack of understanding of app development projects. Furthermore, we assist with a starting point for practitioners to establish hardware and OS-independent guidelines and model understandability for B2X mobile app development and its included stakeholder [20, 88–90].

## Limitations, further research directions, and conclusions

Further in-depth examination of app development processes in practice remains interesting, although our expert interviews generated a detailed and saturated insight into typical app development processes. They can refine the layers presented in REMOB. Also, the relationships between the layers and boundaries of these relationships can be further quantitatively tested. Independent of the background, e.g., technology-, business-, or user-oriented, our interviewees focused on parts of the development process. To explain the integration of theoretical knowledge in practice, direct observations can generate more information about the underlying development processes. In any case, an evaluation must include the different views of relevant stakeholders in the mobile app development process and the usefulness of REMOB. Qualitative [91] or quantitative instruments [92] can serve as efficient reference model evaluation guidelines.

We focused on interviews with European experts; only a few had experiences outside Europe, including the U.S.A., China, and India. Cross-national and cross-cultural research can be useful to understand better development practices in cross-cultural teams, their skills, and development processes, as already marked by some experts, such as app developer E6 and PO E16. Because cultural factors of app usage in different app domains have already been examined [93], this avenue for research can advance the understanding of critical success factors for apps. Cross-national examinations of development processes and their project organization are especially important for (mobile) software developers in exceptional situations like the COVID-19/22 pandemic, where travel restrictions arise.

First, we reviewed the literature on process models for the development of mobile apps. Second, we conducted 28 semi-structured expert interviews with various stakeholders to examine and discuss the state of the art, reasons for (non-)usage of process models, and chances and challenges associated with their usage. In line with the literature, our interviews revealed that agile and iterative development process models, such as Scrum, are mostly preferred. Our interviews further revealed that, in practice, no development process model is used in its pure academic form for mobile app development. Often, Scrum roles are adapted, or additional roles are added, e.g., a lead PO.

Further, strict documentation requirements of WM are softened to enable combined hybrid development process models. Because of the diverse tailoring of process models in practice, we propose REMOB, a reference model for B2X mobile app development that incorporates and describes the specifics of mobile app development on a prescriptive and generalized level. It can serve as a blueprint for efficient B2X app development. Also, REMOB can serve as a starting point for ongoing academic and practical discussions about the importance of crucial factors for B2X app development on a more unified level.

## Acknowledgements and Funding Information

The authors would like to thank all interviewees that participated for this study. Also, the authors would like to thank all involved people that gave feedback on earlier drafts of this manuscript. This research was partly funded by the Lower Saxony Ministry for Science and Culture and Volkswagen Foundation within the PhD program "Design of Mobile Information Systems in the Digital Transformation" under the project number 21-78904-63-9/16.

# Appendix: Holistic B2X Mobile Application Development – A Reference Model

Table A-3: Interview guideline

| Interview block | Questions |
|---|---|
| Introduction | • Can you briefly outline your professional background and what position you currently hold in your company? |
| Process models in general | • Which process models for (mobile) software development do you know?<br>• Which process models do you use for (mobile) software development and why?<br>• Which additional methods, processes or tools would have improved the development?<br>• Which positive/negative influences have the greatest impact on the process model? |
| B2X mobile application development and process models | • Why is it important to develop mobile apps using process models? Why are process models needed?<br>• To what extent do the process models used for mobile apps differ from the process models used for general software systems?<br>• What are their differences or similarities? |
| B2X mobile application domain-specificities | • What is the most important thing to keep in mind when developing such an app (in domain x)?<br>• What do you think are the most important challenges that app developers face when developing a mobile app?<br>• What are their differences in the development of different app domains? E.g., the difference between banking and social media apps? |
| Ending | • Imagine getting up with the right foot tomorrow. Your supervisor comes to you and says you can decide how your team will approach the development of mobile apps in the future. Which suggestion would you make?<br>• In your opinion, are there any important points that we have not addressed? |

Table A-4: First order codes, anchor examples, second order themes, and aggregated dimensions identified

| First order codes and anchor examples | Second order themes | Aggregated dimension |
|---|---|---|
| Statements about the usage of Scrum for the B2X app development – “We work in an agile software project, if you want to call it that, and proceed according to Scrum. We try to orient ourselves as closely as possible to the doctrine, i.e., we try to manage all the events that exist there, from sprint planning to review and the usual stand-ups.” (E5) | Scrum | Selection of a specific process model |
| Statements about the usage of WM for the B2X app development – “It is very important to check whether the requirements are clear. If the requirements are relatively simple and the technology used is relatively clear, then the waterfall model is relatively well suited. We have often recommended and used the waterfall model ourselves.” (E2) | WM | |
| Statements about the usage of XP or SM for the B2X app development – “But in principle, XP came into being in such a way that one tried to summarize this […] and to formalize somehow in my opinion. The normal software approach/ oh I don't know, which naturally arises in such a way, if one tries to stemmed such a project. Well. To do Scrum alone, that makes no sense I think.” (E12) | Others | |
| Statements about adaptions of Scrum for the B2X app development – “Well, you always have to look at whether I can adopt a methodology from a textbook one-to-one, whether it fits into the company. So, and there it is usually not always necessarily the case. For example, we have / we do three-week sprints, I think with Scrum one generally rather assumes two weeks or even shorter, we just go for three-week sprints.” (E10) | Adaptions of Scrum | Hybridization and adaption of process models |
| Statements about sequential and iterative process model combination for the B2X app development – “No, so we also work in the definition phase in the waterfall model and the/ We have after integration phase/ We can run quite flexibly in parallel. […] On the contrary, it is actually the case that two or three banks have now come and said, man, this apparently works so well that they then also have such a Scrum team next to it that the sprints and the scopes are again coordinated in such a way that they also want to go in the direction of Scrum. So that's already the case, and it comes up quite often.” (E16) | Sequential and iterative process model combination | |
| Statements about client or customers' requirements and perspectives in B2X app development processes – “Then I also work a lot with our developers when it comes to building the code in such a way that the requirements are actually implemented, as desired by all those involved, both by all parties involved, both by the potential users and by the clients.” (E21) | Clients / Customers perspective | Requirements engineering for apps |
| Statements about testing in B2X app development processes – “I would say that the tests are looked at more closely when it comes to payments, receivables and so on. They look at what I'm testing there, is it somehow critical or is a button not there at the bottom left? Of course, this has a different priority when testing.” (E5) | Testing | |
| Statements about role allocation of team members in B2X app development processes – “And whether we chose Kanban or Scrum was often based on whether the PO could be named. In other words, whether we could find a contact person who could provide strong technical input. Very often there was a lack of this role, so that in some projects we also started with a Kanban Board and then distributed this expertise to several people.” (E2) | Role allocation | Team characteristics and roles in development processes |
| Statements about lack of understanding of B2X app development processes – “But if you have an experienced team really and out of the team come just from the learning. Scrum is an empirical learning process and then you can but also make good adjustments to it, which also bring many benefits. Inexperienced Scrum masters and inexperienced teams keep then and if they are allowed close to the process, but the more you learn and the more experience you have, the better you can make just also meaningful deviations, which then also really have an added value.” (E28) | Lack of understanding | |
| Statements about the communication between team members in the B2X app development – “In other words, it was very, very good communication that made many things faster and easier. So just such a developer who is technically much deeper in it and then has an idea how you could simplify the process by a lot and who also has experience from other projects.” (E24) | Communication between development team members | Communication and information flows in development processes |
| Statements about communication with stakeholders in the B2X app development – “We do that too, that's one point, that is, our clients of course gets these versions again and again in between, even every few weeks and looks at it and gives us feedback, so that's one point that we simply pull up from agility […].” (E10) | Communication with stakeholders | |
| Statements about performance indicators and risk management in B2X app development processes – "The fact that the developers also have problems is related to the fact that they also have to make a commitment, what am I going to do for the next three weeks, when they are always subject to great uncertainty. […] We are also bound by release dates. We have four main releases a year that we have to stick to, just like all the other departments, and of course that kind of destroys our own planning cycle a bit.” (E5) | Performance indicators and risk management | Governance issues and management of development processes |